\documentclass[nonacm, sigconf]{acmart}
\AtBeginDocument{%
  \providecommand\BibTeX{{%
    \normalfont B\kern-0.5em{\scshape i\kern-0.25em b}\kern-0.8em\TeX}}}

\usepackage{booktabs} % For f\textsl{S}ormal tables
\usepackage[linesnumbered,algoruled,boxed,lined]{algorithm2e}
\usepackage{pdfpages}
\usepackage{graphicx}
\usepackage{textcomp}
\usepackage{xcolor}
\usepackage{adjustbox}
\usepackage{multirow}
\usepackage{booktabs}
\usepackage{rotating}
\usepackage[inline]{enumitem}
\usepackage[para]{threeparttable}
\usepackage{etoolbox}
\usepackage{balance}
\usepackage{microtype}
\usepackage{float}
\usepackage{array}  
\usepackage{balance} 

\begin{document}

\title[A Valid Self-Report is Never Late, Nor is it Early]{A Valid Self-Report is Never Late, Nor is it Early: On Considering the "Right" Temporal Distance for Assessing Emotional Experience}

%AUTHORS
\author{Bernd Dudzik}
\authornote{This is the corresponding author}
\affiliation{%
    \institution{Delft University of Technology}
    \streetaddress{Van Mourik Broekmanweg 6}
    \city{Delft}
    \state{South Holland}
    \postcode{2628 XE}
    \country{The Netherlands}}
\email{B.J.W.Dudzik@tudelft.nl}

\author{Joost Broekens}
\affiliation{%
	\institution{Leiden University}
	\streetaddress{Van Mourik Broekmanweg 6}
	\city{Leiden}
	\state{South Holland}
	\postcode{2333CA}
	\country{The Netherlands}}
\email{d.j.broekens@liacs.leidenuniv.nl}

\renewcommand{\shortauthors}{Dudzik and Broekens}

\begin{abstract}
Developing computational models for automatic affect prediction requires valid self-reports about individuals' emotional interpretations of stimuli. In this article, we highlight the important influence of the temporal distance between a stimulus event and the moment when its experience is reported on the provided information's validity. This influence stems from the \emph{time-dependent} and \emph{time-demanding} nature of the involved cognitive processes. As such, reports can be collected \textit{too late}: forgetting is a widely acknowledged challenge for accurate descriptions of past experience. For this reason, methods striving for assessment as early as possible have become increasingly popular. However, here we argue that collection may also occur \textit{too early}: descriptions about very recent stimuli might be collected before emotional processing has fully converged. Based on these notions, we champion the existence of a temporal distance for each type of stimulus that maximizes the validity of self-reports -- a "right" time. Consequently, we recommend future research to
\begin{enumerate*}
    \item consciously consider the potential influence of temporal distance on affective self-reports when planning data collection,
    \item document the temporal distance of affective self-reports wherever possible as part of corpora for computational modelling, and finally
    \item and explore the effect of temporal distance on self-reports across different types of stimuli   
\end{enumerate*}.

\end{abstract}



% CCS
%\keywords{datasets, neural networks, gaze detection, text tagging}

\maketitle

% ============================================================================
% ============================================================================
\section{Introduction}

A core issue of Affective Computing is developing automatic approaches to predict human emotion by analyzing multimodal sensor data. This enterprise's key resources are corpora of training data, containing examples that match such sensor data with corresponding self-reported subjective experience labels. Supervised datasets of this kind form the basis for teaching computational models to discriminate between different affective states using state-of-the-art machine learning techniques. While other machine perception areas, such as object recognition, can rely on vast pools of relevant training data being available online (e.g., through social media), predicting emotional experience does not allow for the same strategy. Because feelings are highly subjective phenomena, corpora for research often need to be constructed from scratch. Doing so may require involved procedures, whereby individuals are exposed to  conditions approximating those to which responses should eventually be recognized in an application scenario (e.g., in face-to-face conversations \cite{Park2020}, interactions with software applications \cite{Hazer-Rau2020}, or consuming video content \cite{Miranda-Correa2017}). Therefore, the ecological validity of the data collection procedures used for generating training corpora is an important aspect. Without a dataset that is representative of emotions under valid conditions, computational models trained on it will struggle to generalize, potentially failing to deliver accurate predictions when encountering slightly different conditions in the real world.

Emotions are typically distinguished from other types of affective phenomena -- such as moods, or affective attitudes -- by their comparatively short duration and their strong directness towards a triggering stimulus event (see e.g. the taxonomy developed by Scherer \cite{Scherer2005}). In this article, we highlight that one important influence on the validity of any reported affective information in training corpora is the \emph{temporal distance} between the moment at which a self-report is provided and the occurrence of the stimulus event that the report pertains to. While this feature of self-report procedures has been discussed in the literature as a methodological issue for validity, the existing discourse has primarily focused on mitigating inaccuracies related to memory decay over time \cite{Schwarz2007}. However, here we draw on findings from empirical psychology indicating that reports of emotional experience cannot only be collected \textit{too late} and, as a consequence, suffer from memory-biases, but also might be captured \textit{too early}. In this case, self-reports could be requested or provided at moments before individuals' naturally occurring emotional processing of a stimulus has terminated, leading to artificially distorted reports. 

Based on these insights, we suggest choosing the "right" distance for asking individuals about their emotional experience is crucial in designing data collection procedures and should be thoroughly explored in future research endeavours. Concretely, we recommend 
\begin{enumerate*}
    \item conscious consideration of the potential influence exercised by temporal distance on affective self-reports when planning data collection,
    \item documentation of the temporal distance of affective self-reports wherever possible as part of corpora for computational modelling, and finally
    \item a systematic exploration of the influence of temporal distance within and across different types of stimulus events   
\end{enumerate*}.

% Definition of Stimulus Event
% -- Definition of event or stimulus is important
% -- 
% -- There is probably a difference in how
% -- All evaluations are affective attitudes 

% ============================================================================
% ============================================================================
\section{Don't Be Too Late: Avoiding Recall-Biases in Large Temporal Distances}

The temporal distance of self-reports as a methodological issue has been discussed extensively in the social sciences, focusing especially on minimizing potential recall biases resulting from large temporal distances \cite{Schwarz2007, NapaScollon2009}. In this context, strategies for collecting self-reports are often coarsely divided into two categories for discussion, based on the timing at which they prompt individuals \cite{Schwarz2007}: On the one hand \emph{Retrospective Assessments}, which require individuals to reconstruct their experiences from memory at some point in time after a situation. On the other hand \emph{Concurrent Assessments}, which require individuals to immediately (or at least as quickly as possible) report on their feelings. 

Retrospective assessments are a widespread and convenient way for researchers to collect data about individuals' experiences in any survey study. Because these rely solely on participants' memories, questions can target experiences at any point in their past. However, empirical findings demonstrate that details of past experiences quickly become inaccessible as time progresses \cite{Rubin1996}, and recollected memories can be severely distorted as a result (see, e.g., the seminal work of Loftus and colleagues \cite{Loftus1974}). Moreover, biases in autobiographical memory may operate without conscious awareness of the person providing self-reports, resulting in descriptions that tend to unduly correspond to their current beliefs \cite{Levine1997} or other aspects of the situation at recollection \cite{Levine2002}. Due to consistent findings on such recall biases, retrospective self-reports have been deemed to be of questionable validity for collecting information about online experience \cite{Schwarz2007}. 

Because concurrent assessments are taken temporally close to stimulus events of interest, these do not suffer from potential recall biases to the same extent. Perhaps the most common form of concurrent self-reporting used in longitudinal studies are different variants of the \textit{Experience Sampling Method (ESM)} \cite{NapaScollon2009}. In this setting, individuals are typically required to fill in questionnaires about their momentary thoughts or feelings at regular or random time-intervals throughout their daily routine over a prolonged period (i.e., weeks or months), often with the help of technological devices and applications (see Denman et al. \cite{Denman2018} for an example). In addition to time-based schedules for assessments, some studies have also relied on ESM for gathering insights on subjects' impressions about specific stimulus events once these occur \cite{Shiffman2008}. For example, studies have asked individuals to report about recent social interactions \cite{Columbus2020, Dudzik2018episode}. To collect very fine-grained concurrent self-reports throughout specific stimulus events of interest with a short duration (e.g. in the range of seconds to minutes), researchers have furthermore developed a plethora of specialized instruments for collecting time-continuous ratings (e.g. the \textit{Affect Rating Dial} \cite{Ruef2007} or \textit{FeelTrace} \cite{Cowie2000}). Using these instruments, participants provide real-time self-reports while watching video recordings depicting themselves in previous interactions (e.g., Park and colleagues \cite{Park2020}) or rate their experience throughout short-term media content (e.g., in the studies of Mauss et al. \cite{Mauss2005}, or Zhang and colleagues \cite{Zhang2020}). 
    
In summary, concurrent assessments are considered superior to retrospective methods in terms of validity, primarily because they minimize the temporal distance between some stimulus event and the corresponding self-report to rule out recall biases. In doing so, they strive to avoid assessing experiences \emph{too late} at any cost. However, concurrent methods suffer from their own potential drawbacks to validity (see Schwarz \cite{Schwarz2007} for an overview). For example, they can interfere with ongoing tasks, and especially requirements for repeated assessments or strict response times lead to selection biases in participants \cite{NapaScollon2009}. Technological research has attempted to streamline concurrent data collection methods to dampen these adverse effects, e.g., by considering mental workload in developing rating tools \cite{Zhang2020} or automatically determining moments convenient for participants to respond to ESM probes \cite{Mehrotra2015}.

% ============================================================================
% ============================================================================
\section{On being Too Early: Potential Effects of Small Temporal Distances}

Beyond concerns about disturbing or mentally overtaxing individuals when concurrently probing for experiences, minimizing the temporal distance of self-reports may have other negative influences on validity. In this section, we draw on psychology findings to argue that meaningful and stable descriptions of emotional experience may require time to arise and develop in individuals -- and for more complex stimuli more so than for simple ones.         

Cognitive appraisal theories are a prominent conceptualization for the process by which stimuli elicit emotional responses in human beings \cite{Moors2013}, and formed the foundation for many approaches at computationally modelling emotions (see the review of Marsella et al. for an overview \cite{Marsella2010}). They postulate that emotions arise from an organism's cognitive evaluations of its environment regarding its capacity for obstruction or fulfilment of persistent goals and needs \cite{Moors2013}.

Typically, appraisal theories posit a range of variables in terms of which stimuli are evaluated, such as the degree of familiarity to the individual or whether they relate oneself or others \cite{Ortony1990a}. Because appraisal is a subjective interpretation, evaluation of the same stimulus event can result in different outcomes across individuals or the same individual over time (e.g. familiarity might change). While scholars have traditionally focused on exploring the structure of appraisals underlying types of emotional responses (i.e., by identifying the variables necessary to differentiate between different outcomes), research has also developed accounts of the cognitive process underlying appraisal dynamics \cite{Moors2013}. In alignment with theoretical accounts of human cognition more generally (i.e., under the umbrella of dual-process theories \cite{Evans2013}), appraisal theorists envision it as a continuously operating process of varying complexity, taking place at different speeds and degrees of conscious awareness \cite{Clore2008}. Moreover, some theories suggest that various aspects of stimulus evaluations depend on each other, with the outcomes of simple, largely pre-conscious stages of evaluation triggering and shaping higher-level cognitions \cite{Moors2013}. For example, Scherer \cite{Scherer2005} proposes a model where appraisal consists of such sequential stimulus checks increasing in complexity over time and depending on the outcome of prior processing. In this account, early stages act as relevance filters (e.g., checks for sudden changes in the environment) for further processing for their relationship to personal goals and eventually social norms and values. While the dynamics of emotional appraisal processing are the topic of ongoing research \cite{Moors2013}, there is some empirical evidence lending support to both the sequential nature of appraisal processing \cite{Scherer2013}, as well as the capacity for the same appraisals to be computed in a more or less automatic fashion -- e.g. depending on prior familiarity and practice \cite{Moors2010}.

Conceiving of emotional experience as the result of such a continuously ongoing, evolving, and variable appraisal process holds several important insights for understanding the influence of temporal distance on self-reports:

\begin{enumerate}
    \item individuals may continue to emotionally process stimuli over some time at increasing degrees of complexity
    \item when doing so, they continuously integrate new information (perceived or remembered, as well as the outcomes of other appraisals), and
    \item the nature of processing is not universally the same, due to differences in the subjective relevance, prior experience, and inherent complexity of the evaluated stimuli
\end{enumerate}.

These attributes make emotional processing both a \textit{time-dependent} as well as a potentially \textit{time-consuming} affair. Moreover, they underline its context-sensitive nature: precisely what processing has taken place after a fixed amount of time may vary across different types of evaluated stimuli and different individuals doing the evaluation. Together, this points to the very real risk that self-reports with a small temporal distance may arrive \emph{too early} -- before emotional processing has fully converged -- and that there may be no one-size-fits-all minimal temporal distance.

% ============================================================================
% ============================================================================
\section{Summary and Conclusions}

Collecting valid self-reports about individuals' experience is a challenging methodological endeavour but vital for progress on technology for understanding and predicting it in real-world conditions. We have argued that the temporal distance between a stimulus event and a self-report about its experience plays an important role in this regard. One way in which problems for validity may arise is if this distance is too large. Under such circumstances, memory decay may set in and recall biases may manifest themselves,  leading to inaccurate self-reports of experience. This consequence is a widely acknowledged methodological issue, and we have discussed how it is combated by striving for concurrency in assessments. However, drawing on the notion of a dynamic, evolving, and context-sensitive appraisal process underlying emotional responses, we have pointed out that self-reports might similarly arrive too early. While the time-dependency of appraisal processing is still largely unexplored in psychological research, it seems critical for valid data collection. Failing to account for it may result in corpora of idiosyncratic data, capturing instances of incomplete emotional processing that do not generalize to other populations or similar stimuli. Doing so requires collecting self-reports at the "right" time -- a sweet spot in terms of temporal distance that is close enough to minimize the influence of memory decay while also providing enough space to let emotional processing play out naturally. 

 % ============
% FIGURE -- EmoTime_MEEC_Tradeoff
\vspace{1em}
\begin{figure}[!htb]
    \includegraphics[width=\columnwidth]{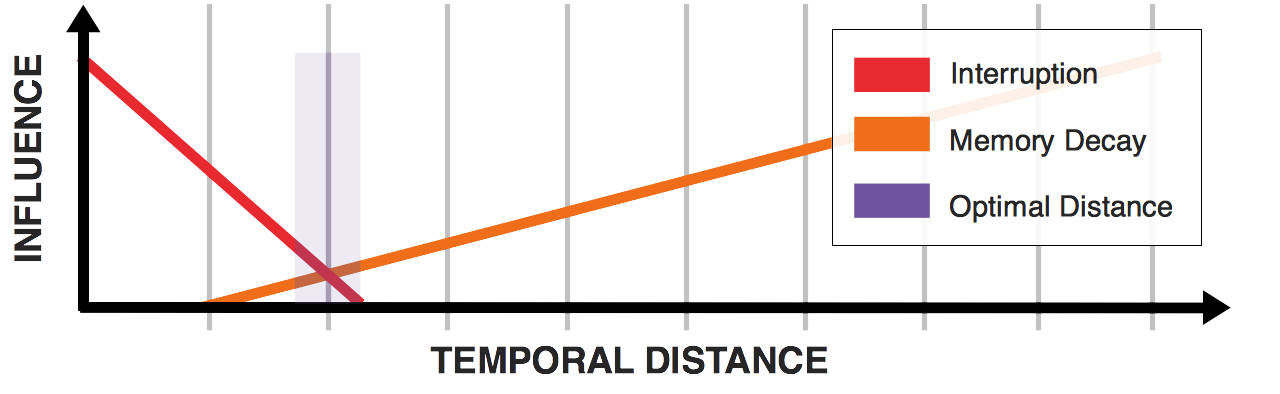}
    \vspace{-1em}
    \caption{Schematic depiction of the optimal temporal distance for assessment after a stimulus event, minimizing both the potential influence of interrupting emotional processing before convergence (too early), as well as decaying memory fidelity (too late) on self-reported emotional experience.}
    \label{fig:EmoTime_MEEC_Tradeoff}
\end{figure}
% % ============

We present a schematic depiction of this idea in \textit{Figure \ref{fig:EmoTime_MEEC_Tradeoff}}. Note that this graph serves a purely illustrative purpose: to the best of our knowledge, how the magnitude of these two influences change over time is currently not known and likely varies across stimuli and the individuals interpreting them. However, it seems reasonable to assume that emotional interpretation may -- at least in some instances -- still be ongoing by the time the effects of memory decay on self-reports start. Evidence for this line of reasoning comes from the overall rapid decline of memory \cite{Rubin1996}, and studies that show differences already after a short time for self-reported experiences of comparatively simple stimuli (i.e., throughout a painful medical procedure \cite{Redelmeier1996}). However, given the postulated evolutionary roots of emotions as rapid response mechanisms for adaptive behavior postulated by some affective scientists (see e.g. the arguments of Levenson \cite{Levenson2003}), it seems likely that emotional processing for many types stimulus events will terminate rather quickly. Thus, while completely concurrent assessments (i.e., with a temporal distance close to zero) may run a substantial risk of arriving before emotional processing has converged on an outcome, assessments with even moderate delays might suffice for simple stimulus events. Given that delays up to this range are already an established practice in the design of existing ESM studies for pragmatic reasons \cite{NapaScollon2009}, these may be less susceptible to too early assessment. However, without explicitly acknowledging the potential role of temporal distance for valid data collection and conducting further empirical exploration on the topic, a fine-grained understanding of the temporal dynamics between interrupted processing and memory decay remains out of reach. 

For this reason, we urge researchers that collect data of self-reported emotional experiences to consciously consider the potential influence of temporal distance in their designs. In particular, for complex stimulus events (e.g., ones touching upon social norms and values), it might be beneficial to combine any concurrent assessments with a form of retrospective self-report and explore potential differences in analysis and modelling. Moreover, it seems important to consider the temporal duration of stimulus events themselves as another potential aspect of their complexity that might influence how they are processed in assessment. Consequently, experimental setups for data collection should carefully choose and document the duration of stimulus events used for elicitation. While capturing the duration of stimulus events in assessments via ESM more difficult, collecting self-reports about duration when probing may at least provide a coarse indication and will help to control for its influence in modeling. 

Similarly, we recommend that researchers document the temporal distance between a stimulus event and the moment when affective self-reports are provided wherever possible. Such documentation should also be undertaken by studies following a cued-recall protocol for data collection where individuals reconstruct their feelings in a past situation while annotating video footage of it with continuous ratings (see the studies of Mauss et al. \cite{Mauss2005}, or Park and colleagues \cite{Park2020} for examples). 

Finally, we urge researchers to systematically explore the effect of temporal distance on self-reported affective experience in dedicated empirical investigations to identify the right time for assessment. Such studies would require repeated assessments of experience at different temporal distances and exploring differences across individuals with different backgrounds and stimuli of different types. Variables that are potentially relevant for shaping both emotional and memory processing (e.g., goals and personal values \cite{Scherer2005, Levine1997}, or mood \cite{Levine2002}) make particularly promising candidates in that respect.

\begin{acks}
    This research was (partially) funded by the Hybrid Intelligence Center, a 10-year programme funded by the Dutch Ministry of Education, Culture and Science through the Netherlands Organisation for Scientific Research, \url{https://hybrid-intelligence-centre.nl}.
\end{acks}

\balance

% LITERATURE LIST
% ========================================================================= 

\bibliographystyle{ACM-Reference-Format}
\bibliography{./EmoTime_arXiv.bib}

\end{document}